\documentclass[12pt]{article}
\usepackage{fullpage}[letter] % %

\usepackage{amsmath, amssymb}
\usepackage{graphicx}
\usepackage{subcaption}
\usepackage{hyperref}
\usepackage{makecell}
\usepackage{array}

\usepackage{msc}

\graphicspath{ {./images/} } %Importing a figure

\begin{document}

\title{An Extended Consent-Based Access Control Framework: Pre-Commit Validation and Emergency Access}

\author{
Nasif Muslim and Jean-Charles Grégoire\\
\textit{Énergie Matériaux Télécommunications Research}\\
\textit{Institut national de la recherche scientifique (INRS)}\\
Montréal, Canada\\
Nasif.Muslim@inrs.ca, jean-charles.gregoire@inrs.ca
}

\date{}
\maketitle

\begin{abstract}
Consent-Based Access Control (CBAC) is a foundational mechanism for enforcing
patient autonomy in modern healthcare information systems. Many CBAC frameworks
are built on the eXtensible Access Control Markup Language (XACML) and inherit
its \emph{lazy evaluation} model, in which policy interactions are resolved only
at request time. This design allows contradictory consent directives to accumulate within the repository, creating a semantic gap between patient intent and system behavior while burdening high-frequency runtime decisions with complex conflict resolution.
This paper presents an extended CBAC framework that enforces semantic
correctness at consent creation time rather than during access evaluation. The
framework introduces a pre-commit validation workflow centered on a Consent
Conflict Analysis Module (CCAM), which proactively detects modality conflicts
and redundancies before directives become active. In addition, immutable system
invariants are formalized to guarantee baseline access for record authors and
patients, preserving clinical continuity and professional accountability.
Finally, the framework incorporates a context-aware emergency mediation
mechanism that enables controlled \emph{break-the-glass} access driven by
real-time physiological evidence, with disclosure strictly bounded by an
Emergency Disclosure Control Function (EDCF).
Simulation-based evaluation using controlled synthetic workloads demonstrates
that pre-commit conflict resolution yields low and stable runtime decision
latency and consistently outperforms standard XACML-based baselines as policy
repositories scale. Emergency access experiments further demonstrate strong restriction to data access, pruning the majority of non-relevant record elements while
preserving clinically essential information.
\end{abstract}

\section{Introduction} \label{sec:introduction}

Modern healthcare systems generate and manage longitudinal health records that
span multiple clinical episodes, healthcare professionals, and healthcare
institutions. In this complex environment, patient consent is not merely
optional: it is a fundamental legal and ethical requirement that must be
enforced at scale. However, emergencies introduce unavoidable exceptions, as
delaying access to health records in time-critical situations can directly endanger patients safety. Consent management in healthcare is, therefore, defined by a persistent tension between \emph{patient autonomy} and \emph{clinical continuity}.

Traditional access control models are poorly suited to these competing demands.
Role-Based Access Control (RBAC), for example, is inherently
administrator-centric and cannot readily accommodate individualized patient
preferences at scale. Although Attribute-Based Access Control (ABAC) offers the
expressivity required for fine-grained authorization, effective consent
enforcement in healthcare requires additional capabilities. These include
explicit lifecycle semantics encompassing creation, activation, revocation, and
expiration, as well as systematic mechanisms for resolving conflicts among
overlapping consent directives. In practice, many healthcare consent-enforcement
systems realize these capabilities by layering consent semantics on top of
general-purpose ABAC infrastructures. 

Many Consent-Based Access Control (CBAC) frameworks are built on the eXtensible Access Control Markup Language (XACML) architecture~\cite{oasisXACML3}. A defining feature of XACML is its runtime policy evaluation model, in which policy conflicts and interactions are resolved only when an access request reaches the Policy Decision Point (PDP) (see Section~\ref{subsec:xacml-architecture}), rather than being validated at policy creation time, a behavior commonly characterized as \emph{lazy policy evaluation}~\cite{Sauwens2021}. Overlapping or contradictory directives are not analyzed when admitted to the policy repository; instead, they are arbitrated on the fly using policy-combining algorithms such as Deny-Overrides. As a result, system robustness depends on runtime arbitration rather than on the proactive maintenance of a semantically consistent consent state.

This reliance on lazy evaluation introduces two critical limitations in
healthcare settings. First, XACML provides no feedback to policy authors
regarding how a newly submitted directive will interact with existing policies.
Consequently, non-expert users (e.g., patients) may create directives that are
syntactically valid but semantically ineffective. Fisler et
al.~\cite{fisler2005verification} characterize this discrepancy as a
\emph{semantic gap} between intended policy meaning and enforced behavior. For
example, a patient attempting to deny access to external specialists by
targeting a department attribute may inadvertently leave the restriction
unenforced due to attribute-matching and policy precedence rules.

Second, the absence of upfront validation allows contradictory directives to
coexist within the policy repository. Over time, this leads to latent anomalies
such as shadowing, redundancy, and direct conflicts, which remain undetected
until triggered by a specific access request. At that point, policy-combining
algorithms produce a decision without resolving the underlying inconsistency.
Hu et al.~\cite{hu2013discovery} show that such anomalies are prevalent in
unvalidated XACML policy sets.

Beyond the limitations of lazy evaluation, existing CBAC frameworks leave two further challenges insufficiently addressed~\cite{zhang2016consent,
peyrone2022formal,jaiman2020consent}. First, most frameworks adopt a
deny-by-default evaluation model~\cite{mousaid2020toward}, failing to formalize baseline access guarantees for record authors and patients themselves, an omission that can undermine clinical continuity and professional accountability. Second, strict consent enforcement alone is inadequate in safety-critical environments. During emergencies, access decisions must incorporate real-time clinical context while preventing unrestricted disclosure and ensuring post-hoc accountability.

Recent advances in Large Language Models (LLMs) have led to their increasing
adoption as reasoning interfaces for authoring, interpreting, and analyzing
access control policies~\cite{vatsa2025exploring,Jayasundara2024}. Prior work
shows that LLMs can translate consent directives expressed in natural language
into formal access control policies, lowering the barrier for non-expert users
(e.g., patients) to articulate complex authorization intent~\cite{Paratore2025,Lawal2024}.
Importantly, these approaches preserve deterministic enforcement at the PDP by
positioning LLMs upstream in the policy lifecycle rather than as decision-making
components.

However, LLM-based interfaces are inherently probabilistic and susceptible to
semantic ambiguity, misinterpretation, and manipulation via adversarial inputs.
As a result, introducing an LLM-assisted authoring interface adds a new source
of semantic risk: logically well-formed but with unintended effect directives
may be admitted into the policy repository and then enforced with technical
precision by the PDP, exacerbating the semantic gap between intended and
enforced behavior.

To address these limitations, this paper presents an extended
CBAC framework that enforces semantic alignment between
patient intent and system enforcement. The proposed approach resolves policy
anomalies at commit time while formalizing default access behavior and
emergency safety mechanisms. The main contributions of this work are as
follows:

\begin{enumerate}
    \item \textbf{Pre-Commit Consent Validation:}
     An extended XACML-based architecture is introduced that incorporates a Consent Conflict Analysis Module (CCAM) (see Section~\ref{subsec:architectural_components}) to detect and eliminate policy anomalies such as access-mode conflicts and redundancies before directives become active.

    \item \textbf{Formal System Invariants:}
    Immutable authorization invariants are formally defined (see Section~\ref{subsec:system_invariants}) to guarantee
    baseline access for record authors ($\mathit{Author}(r)$) and patients
    ($\mathit{Subject}(r)$), preserving clinical continuity independently of
    dynamic consent changes.

    \item \textbf{Context-Aware Emergency Authorization:}
    A privacy-preserving emergency override mechanism is developed that
    leverages real-time physiological observations from medical Internet of
    Things (IoT) devices to issue context-bound authorizations limited to a
    condition-relevant Clinical Brief (see Section~\ref{sec:emergency_mediator_framework}).
\end{enumerate}

The remainder of this paper is organized as follows. Section~\ref{sec:background}
reviews related work in CBAC and summarizes the XACML reference architecture and
its limitations in healthcare settings. Section~\ref{sec:cbac_framework}
presents the design of the proposed CBAC framework, including its formal data
model and system architecture. Section~\ref{sec:consent_policy_setup}
introduces the consent policy setup workflow and the pre-commit validation
mechanism used to eliminate policy anomalies prior to activation.
Section~\ref{sec:emergency_mediator_framework} describes the emergency override
mechanism and its privacy-preserving, context-aware disclosure logic.
Section~\ref{sec:performance_measurement} evaluates the framework through
controlled performance measurements. Finally, Section~\ref{sec:discussion}
discusses implications, limitations, and directions for future work.

\section{Background}
\label{sec:background}

This section provides background to contextualize the proposed CBAC framework.
Section~\ref{subsec:soa-cbac} surveys state-of-the-art CBAC approaches and their
limitations in healthcare environments. Section~\ref{subsec:xacml-architecture}
then reviews the XACML reference architecture and request-evaluation workflow,
which underpin many practical CBAC implementations and motivate the
architectural extensions introduced later in this paper.

\subsection{State of the Art in Consent-Based Access Control}
\label{subsec:soa-cbac}

Recent research in CBAC has explored diverse
enforcement strategies, including cryptographic mechanisms, formal policy
models, blockchain-based ledgers, and interoperability standards such as
Fast Healthcare Interoperability Resources (FHIR)~\cite{hl7_fhir}.

Zhang et al.~\cite{zhang2016consent} propose an early cryptographic CBAC
approach using conditional proxy re-encryption. Patient records are encrypted
and outsourced to a semi-trusted data center; consent is granted via
re-encryption keys. The method enforces consent cryptographically but supports
only record-level granularity without contextual or attribute-based control.

Peyrone et al.~\cite{peyrone2022formal} present a formal Event-B model that
covers the full consent lifecycle (creation, renewal, withdrawal) and verifies
state consistency. However, the model is limited to specification and offers no
runtime enforcement or conflict resolution for overlapping directives.

Jaiman et al.~\cite{jaiman2020consent} introduce a blockchain-based CBAC
framework using the Data Use Ontology (DUO) and smart contracts for dynamic,
auditable consent management. Enforcement relies on semantic matching of
intended use, yet granularity remains at the dataset level. Moreover, the
framework does not provide an explicit mechanism for interoperable consent
enforcement across heterogeneous systems.

Mousaid~\cite{mousaid2020toward} proposes Consent-Centric Attribute-Based Access
Control (C-ABAC), combining FHIR consent resources with XACML-based enforcement
for fine-grained, interoperable policies. However, the framework inherits
XACML's lazy evaluation limitations, uses a deny-by-default model (which can
deny access even to record authors unless explicit consent is present), and does not support break-the-glass access for emergencies.

Kanwal et al.~\cite{kanwal2019privacy} propose a relationship-semantics-based
XACML model for electronic health records in hybrid cloud settings, enabling
fine-grained, privacy-aware access control based on inter-entity
relationships. However, the model does not address consent lifecycle
conflicts, proactive validation, or emergency override semantics and relies on
runtime policy evaluation.

Table~\ref{tab:cbac_comparison} compares the discussed CBAC models with respect
to key consent enforcement dimensions: default access policy, lifecycle
support, granularity, interoperability, conflict resolution, emergency
access, and auditability.

\begin{table}[t]
\centering
\small
\setlength{\tabcolsep}{3pt}
\caption{Comparison of state-of-the-art consent-centric and XACML-based access
control models for healthcare}
\label{tab:cbac_comparison}
\renewcommand{\arraystretch}{1.2}
\begin{tabular}{|
>{\centering\arraybackslash}p{1.6cm}|
>{\centering\arraybackslash}p{1.5cm}|
>{\centering\arraybackslash}p{1.7cm}|
>{\centering\arraybackslash}p{1.6cm}|
>{\centering\arraybackslash}p{2.0cm}|
>{\centering\arraybackslash}p{2.1cm}|
>{\centering\arraybackslash}p{2.0cm}|
>{\centering\arraybackslash}p{1.6cm}|
}
\hline
\textbf{Model} &
\makecell{\textbf{Default}\\\textbf{Policy}} &
\makecell{\textbf{Lifecycle}\\\textbf{Support}} &
\makecell{\textbf{Gran-}\\\textbf{ularity}} &
\makecell{\textbf{Interoper-}\\\textbf{ability}} &
\makecell{\textbf{Conflict}\\\textbf{Resolution}} &
\makecell{\textbf{Emergency}\\\textbf{Access}} &
\makecell{\textbf{Audita-}\\\textbf{bility}} \\
\hline
Zhang et al.~\cite{zhang2016consent} & Not spec. & Limited & Record-level & No & No & No & Limited \\
\hline
Peyrone et al.~\cite{peyrone2022formal} & Not spec. & Yes & N/A & No & No & No & Yes \\
\hline
Jaiman et al.~\cite{jaiman2020consent} & Not spec. & Yes & Dataset-level & No & No & No & Yes \\
\hline
Mousaid~\cite{mousaid2020toward} & Deny-default & Yes & Attribute-level & Yes (FHIR) & Runtime (XACML) & No & Yes \\
\hline
Kanwal et al.~\cite{kanwal2019privacy} & Deny-default & No & Attribute-level & Yes (XACML) & Runtime (XACML) & No & Yes \\
\hline
\end{tabular}
\end{table}

As highlighted in Table~\ref{tab:cbac_comparison}, no existing CBAC model
simultaneously addresses conflict resolution, emergency overrides, and default
access policies. These limitations motivate the design of the proposed
framework, which extends existing CBAC architectures with proactive policy
validation and context-aware emergency enforcement.

\subsection{XACML Reference Architecture and its Workflow}
\label{subsec:xacml-architecture}

While Section~\ref{subsec:soa-cbac} surveys enforcement approaches for
CBAC, practical implementations often build on
the XACML reference architecture.
XACML provides a modular framework that separates policy administration,
request interception, attribute resolution, and decision evaluation, making it
a common foundation for CBAC systems in healthcare.

In the XACML reference architecture, access control is organized around four
components:

\begin{itemize}
    \item \textbf{Policy Enforcement Point (PEP)}: Acts as the gatekeeper. It
    intercepts access requests, formats them for evaluation, and enforces the
    returned decision (e.g., allowing or denying access to a patient's record).

    \item \textbf{Policy Decision Point (PDP)}: The core evaluator. It retrieves
    policies from the PAP and assesses requests against them, incorporating attributes from the PIP.

    \item \textbf{Policy Administration Point (PAP)}: Manages the policy
    lifecycle. It enables authorized users to create, modify, and store consent
    directives in a policy repository.

    \item \textbf{Policy Information Point (PIP)}: Supplies contextual
    attributes (e.g., user roles, record metadata, or consent status) during
    policy evaluation.
\end{itemize}

\begin{figure}[t]
\centering
\includegraphics[width=15cm]{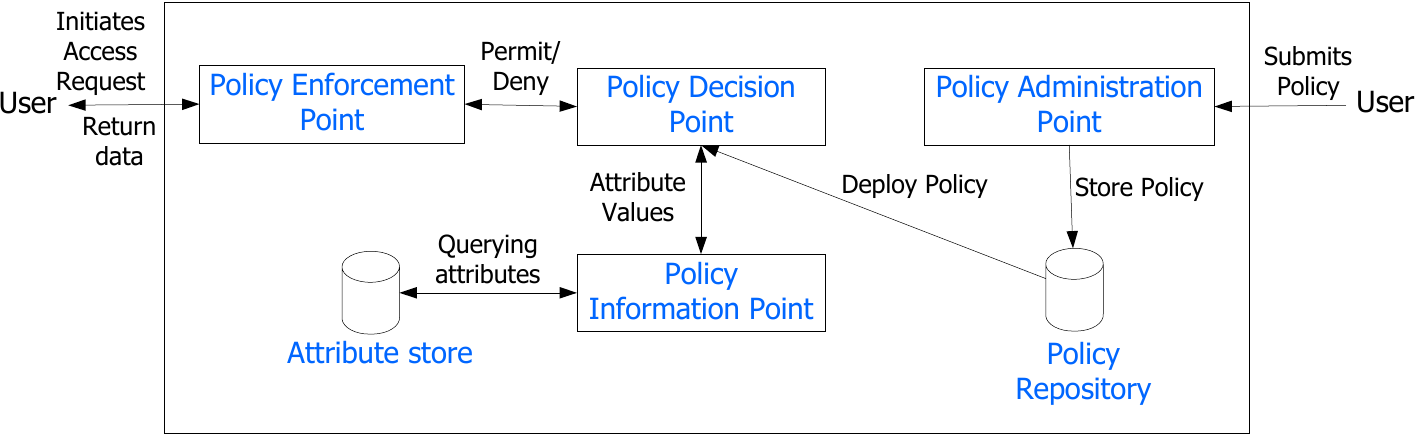}
\caption{Overview of the standard XACML architecture and request evaluation
workflow~\cite{oasisXACML3}}
\label{fig:xacml-architecture}
\end{figure}

In a typical XACML request-evaluation workflow (shown in
Figure~\ref{fig:xacml-architecture}):

\begin{enumerate}
    \item A user (e.g., a healthcare professional) initiates an access request
    for a protected record.
    \item The PEP intercepts the request and forwards it to the PDP.
    \item The PDP retrieves relevant policies from the PAP.
    \item If required, the PDP queries the PIP for additional attributes (e.g.,
    real-time consent status).
    \item The PDP evaluates the request and returns a decision
    (\texttt{Permit} or \texttt{Deny}) to the PEP.
    \item The PEP enforces the decision by granting or denying access.
\end{enumerate}

The modular design of the XACML reference architecture enables fine-grained
policy enforcement but also relies on \emph{lazy policy evaluation}, in which
policy interactions and conflicts are typically resolved only at request time.
This limitation motivates extending the architecture with a pre-commit
validation workflow and related components.

\section{Design of the Consent-Based Access Control Framework}
\label{sec:cbac_framework}

This section presents the design of the proposed CBAC framework, including its
formal data model, architectural components, and lifecycle-aware consent
management. The design establishes the system entities and trust assumptions
required to support predictable consent handling under both routine access and
emergency override scenarios.

The section is organized as follows.
Section~\ref{subsec:formal_data_model} introduces the formal data model that
defines the core clinical entities, consent directives, and access request
structures. Section~\ref{subsec:architectural_components} presents the
architectural components that realize consent-based authorization within an
extended XACML framework. Section~\ref{subsec:consent_lifecycle_management}
formalizes the consent directive lifecycle from creation to termination.
Finally, Section~\ref{subsec:operational_environment-trust-entities} describes
the operational environment and trust entities that contextualize deployment
assumptions and emergency authorization.

\subsection{Formal Data Model}
\label{subsec:formal_data_model}

The CBAC framework is built upon a formal data model of clinical entities and
records, independent of their physical storage in any particular health record
repository system. This data model defines the logical structure of clinical entities, consent directives, and access requests that underpin all authorization
decisions in the system.
The remainder of this subsection formalizes (i) clinical entities and provenance
relationships (Section~\ref{subsubsec:clinical_entities}), (ii) consent directives
defined over these entities (Section~\ref{subsubsec:consent_directive_formalization}),
and (iii) the access request structures (standard and emergency) evaluated by the
authorization framework (Section~\ref{subsubsec:access_request_formalization}).

\subsubsection{Clinical Entities and Provenance}
\label{subsubsec:clinical_entities}

This subsection defines the \emph{healthcare data model}, which provides an
abstract, system-independent representation of clinical entities and their
provenance relationships. These logical entities are derived from, but distinct
from, their physical representation in the underlying health record database.

Let $U$ denote the set of all system users, partitioned into healthcare
professionals $HP \subseteq U$ and patients $Pt \subseteq U$. Let $R$ denote the set of all atomic
clinical records represented in the healthcare data model (e.g., clinical
notes, prescriptions, and laboratory results).

Clinical information is organized hierarchically through the following
relational entities:

\begin{itemize}
    \item \textbf{Episodes ($EP$):}  
    Each episode $e \in EP$ corresponds to a distinct clinical context, such as
    an outpatient visit or a hospital admission. Formally, an episode links exactly one supervising healthcare professional $\mathit{Creator}_e \in HP$ to exactly one patient
    $\mathit{Patient}_e \in Pt$.

    \item \textbf{Clinical Records ($R$):}  
    Each clinical record $r \in R$ is associated with exactly one episode.
    A total function captures this association: $\mathit{ep} : R \rightarrow EP$ meaning that for every record $r \in R$, $\mathit{ep}(r) \in EP$.
\end{itemize}

To support authorization reasoning, the data model exposes the following helper
functions that derive logical provenance attributes from stored health record metadata:

\begin{itemize}
    \item \textbf{$\mathit{Author}(r) : R \rightarrow HP$} \\
    Returns the healthcare professional responsible for creating the record $r$,
    derived from the supervising professional of the enclosing episode.

    \item \textbf{$\mathit{Subject}(r) : R \rightarrow Pt$} \\
    Returns the patient to whom the record $r$ pertains, as determined by the
    episode in which the record was generated.
\end{itemize}

\subsubsection{Consent Directives Formalization}
\label{subsubsec:consent_directive_formalization}

Consent directives are represented as abstract, structured policy objects
that capture patient authorization intent over the healthcare data model
defined above. These directives are logical entities evaluated by the authorization framework and are distinct from their concrete representation in policy repositories or enforcement engines.

Formally, a consent directive $C$ is represented as a tuple:
\[
    C = \langle \mathit{grantee}, \mathit{target}, \mathit{effect} \rangle
\]
where:

\begin{itemize}
    \item \textbf{$\mathit{grantee} \in HP$}:  
    The identifier of the healthcare professional to whom the directive
    applies.

    \item \textbf{$\mathit{target}$}:  
    The authorization scope of the directive, expressed over healthcare data
    model entities. The target may reference a clinical episode
    $e \in EP$ or a specific clinical record $r \in R$, using the
    provenance relationships defined in Section~\ref{subsubsec:clinical_entities}.

    \item \textbf{$\mathit{effect}$}:  
    The access-mode specified by the directive indicates whether access
    to the target is explicitly granted or restricted (\texttt{Permit}, \texttt{Deny}).
\end{itemize}

\subsubsection{Access Requests Formalization}
\label{subsubsec:access_request_formalization}

The proposed CBAC framework supports two distinct types of access requests:
standard consent-based authorization requests and emergency override
authorization requests. Standard requests are target-driven, where a requester
seeks access to a specific clinical record item. In contrast, emergency override
requests are evidence-driven, initiated not by a specific record target but by a
set of authenticated biometric and physiological observations $\mathsf{bioObs}$
that justify broad access, with the scope of disclosure dynamically derived from
clinical relevance.

Access requests are initially constructed using system-level authentication
artifacts supplied by the HRR controller (see
Section~\ref{subsec:operational_environment-trust-entities}) and the underlying
identity infrastructure. Prior to authorization evaluation, these artifacts are
resolved into logical subjects defined in the healthcare data model.

\begin{itemize}
    \item \textbf{Standard (Target-Driven) Access Request:}  
    A standard access request $req_{std} \in Q_{std}$ is defined as:
    \[
        req_{std} =
        \langle
        \texttt{Standard},
        \mathit{Requester},
        \mathit{PatientID},
        target,
        ts
        \rangle
    \]
    where:
    \begin{itemize}
        \item $\mathit{Requester} \in U$: a user (healthcare professional or patient) initiating the access request
        \item $\mathit{PatientID} \in Pt$: the patient identifier associated with the requested clinical data
        \item $target \in R_{item} \cup R_{ep}$: the clinical episode or record being accessed
        \item $ts$: request timestamp
    \end{itemize}

    \item \textbf{Emergency (Evidence-Driven) Access Request:}  
    An emergency access request $req_{em} \in Q_{em}$ is defined as:
    \[
        req_{em} =
        \langle
        \texttt{Emergency},
        \mathit{Requester},
        \mathit{PatientIDClaim},
        \mathsf{bioObs},
        ts
        \rangle
    \]
    where $Q_{em}$ denotes the set of emergency access requests, and:
    \begin{itemize}
        \item $\mathit{Requester} \in HP$: a healthcare professional initiating the access request
        \item $\mathit{PatientIDClaim}
        = \langle bio, \mathit{PatientID} \rangle$: a biometric-backed patient identity claim supplied with the emergency request
        \item $\mathsf{bioObs}$: verifiable biometric and physiological observations used to infer the emergency clinical state (e.g., heart rate, $SpO_2$, blood pressure)
        \item $ts$: request timestamp
    \end{itemize}
    
\end{itemize}

The access request structures defined above serve as inputs to distinct
authorization workflows within the CBAC framework. Standard (target-driven)
requests are evaluated according to the pre-commit validation and authorization
semantics defined in Section~\ref{sec:consent_policy_setup}, whereas Emergency
(evidence-driven) requests are processed through the emergency override
mechanism described in Section~\ref{sec:emergency_mediator_framework}.

Throughout this paper, symbols such as $\mathit{PatientID}$, $EP$, and $R$
denote logical entities in the healthcare data model, whereas monospace
identifiers (e.g., \texttt{patient\_id}, \texttt{requester\_id}) refer to
implementation-level attributes in the underlying health record database used
solely to resolve these logical identifiers.

\subsection{Architectural Components} \label{subsec:architectural_components}

The proposed framework extends the standard XACML reference architecture to
support two primary workflows: Consent Policy Setup and Consent Policy
Enforcement. To achieve this, the architecture replaces generic XACML
components with specialized modules designed to handle clinical provenance
and consent-driven authorization logic.

\begin{itemize}

    \item \textbf{Consent Policy Administration Point (CPAP):} 
    The interface through which patients submit draft consent directives.
    The CPAP performs preliminary formatting and forwards draft directives
    to the CCAM for validation prior to repository admission.

    \item \textbf{Consent Conflict Analysis Module (CCAM):} 
    The component responsible for detecting policy anomalies. It ensures
    that newly submitted directives do not violate system invariants or
    conflict with existing active directives before being stored in the
    repository.

    \item \textbf{Consent Policy Repository (CPR):} 
    The persistent store for validated consent directives. The CPR
    maintains directive states (Draft, Active, Inactive) in accordance with
    the defined consent lifecycle model.

    \item \textbf{Consent Policy Decision Point (CPDP):} 
    The component responsible for evaluating Standard (Target-Driven)
    access requests. Upon receiving a request from the C-PEP, the CPDP
    coordinates with the CPIP to resolve record metadata and with the CPR
    to retrieve active consent directives. It computes authorization
    outcomes by executing a layered evaluation that combines immutable
    system invariants with patient-delegated consent directives. The
    resulting \texttt{Permit} or \texttt{Deny} decision is returned to the
    C-PEP for enforcement.
    
    \item \textbf{Consent-Aware Policy Enforcement Point (C-PEP):}
    The ingress component that receives access requests from healthcare
    professionals and patients. Upon receipt of a request tuple
    ($req_{std}$ or $req_{em}$), the C-PEP extracts the request type to
    determine the appropriate evaluation path.
    For Standard requests, the C-PEP routes the request to the CPDP for
    deterministic policy evaluation against the consent repository. For
    Emergency requests, the C-PEP routes the request to the ECDM for
    context-aware override processing and disclosure control. The C-PEP
    enforces the authorization outcome by either permitting or denying
    access to the target record, or by delivering a condition-relevant
    Clinical Brief during emergency access.

    \item \textbf{Emergency Context and Disclosure Manager (ECDM):}
    A specialized authorization component responsible for coordinating
    Emergency (Evidence-Driven) access requests. It evaluates authenticated
    biometric and physiological observations $\mathsf{bioObs}$ to
    determine the emergency clinical context, derives a semantic emergency
    disclosure scope, and acquires Emergency Override Tokens (EOTs) from
    the Emergency Authorization Authority (EAA). The issued EOT is subsequently used by downstream components (e.g., the HRR) to enable scope-bounded retrieval of clinical episodes and records (Section~\ref{subsec:ecdm}).
    
    \item \textbf{Consent Policy Information Point (CPIP):} 
    The information interface between the authorization components of the
    CBAC framework and the health record database. The CPIP queries health
    record metadata and executes auxiliary provenance-resolution routines
    to derive clinical attributes (e.g., record authorship, subject
    association, and episode context) required for consent-based
    evaluation.

\end{itemize}

\subsection{Consent Lifecycle Management} \label{subsec:consent_lifecycle_management}

Consent directives in the framework are represented as structured policy
objects whose applicability is governed by explicit lifecycle semantics.
To ensure predictable, verifiable behavior during policy evaluation, each directive follows a Finite State Machine (FSM) model that defines its progression from creation through enforcement to termination.

A consent directive can exist in one of three primary states:

\begin{itemize}

    \item \textbf{Draft:} The initial state in which the patient formulates a consent directive. Draft directives are non-enforceable and do not participate in access-control decisions. A directive may transition to the \textit{Active} state only after successful validation by the CCAM.
  
    \item \textbf{Active:} The enforceable state of a consent directive. In this state, the directive is stored in the CPR and evaluated by the CPDP during the enforcement workflow to extend or restrict access rights. An active directive remains in force until the patient explicitly revokes it or it expires upon the conclusion of its validity interval.

    \item \textbf{Inactive:} The terminal state of a consent directive. Once inactive, a directive is excluded from all subsequent CPDP policy evaluations. A directive enters this state through patient-initiated revocation or automatic expiration when the predefined validity period lapses.

\end{itemize}

\begin{figure}[h!]
    \centering
    \includegraphics[width=10cm]{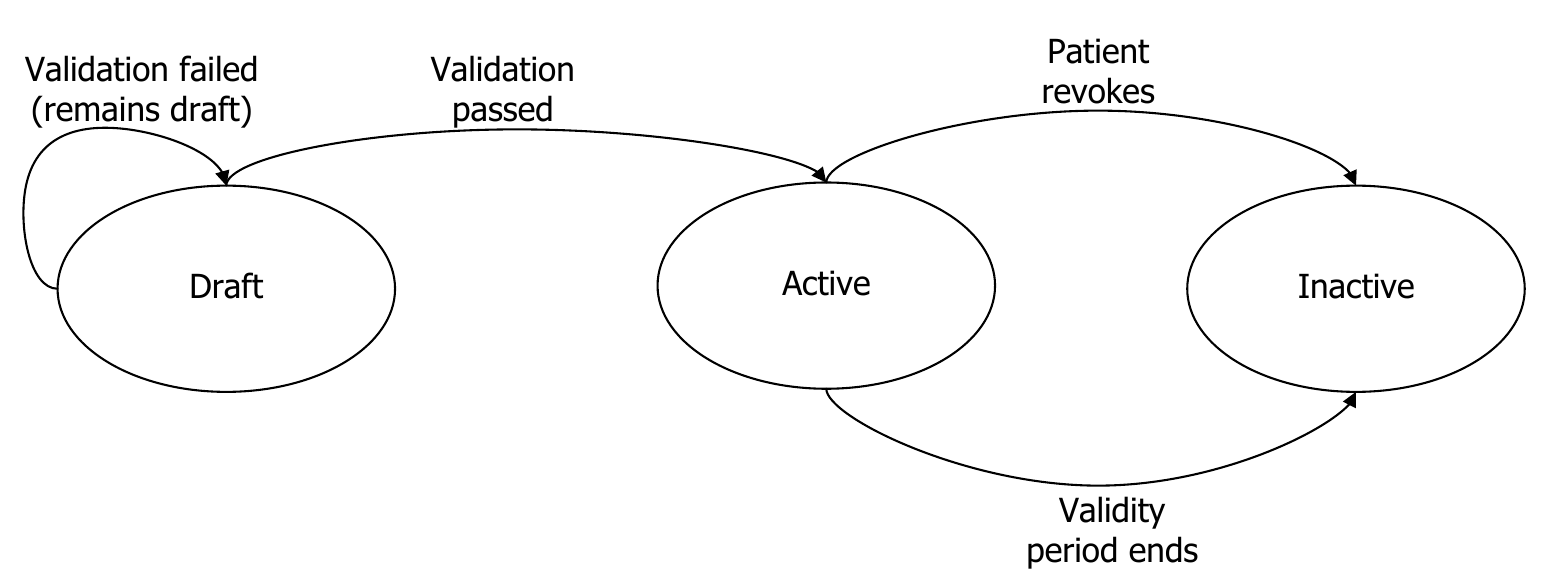}
    \caption{Finite State Machine (FSM) for the consent lifecycle, showing transitions between Draft, Active, and Inactive states}
    \label{fig:consent_lifecycle}
\end{figure}

The lifecycle of a consent directive, including its transitions, is shown in Fig.~\ref{fig:consent_lifecycle}.
Transitions between states are triggered by system events (e.g., validation outcomes, temporal constraints) and patient interventions, as described below:

\begin{itemize}

  \item \textbf{Activation (Draft $\rightarrow$ Active):}
  A draft directive becomes active only when the CCAM verifies that it conflicts neither with system invariants (e.g., authorship rules) nor with any existing active directives. Formally, activation is contingent upon the conflict predicate evaluating to false:
  \[
    \text{Conflict}(C_{\text{draft}}) = \mathsf{False}
  \]

  \item \textbf{Revocation (Active $\rightarrow$ Inactive):}
  A patient may revoke an active consent directive at any time. Revocation has an immediate effect: the directive’s status is updated to inactive within the CPR, ensuring it is excluded from all subsequent evaluations.

  \item \textbf{Expiration (Active $\rightarrow$ Inactive):}
  Each consent directive may optionally include a temporal validity interval. When the current system time exceeds this interval, the directive automatically transitions to an inactive status in the CPR.

\end{itemize}

\subsection{Operational Environment and Trust Entities}
\label{subsec:operational_environment-trust-entities}

To facilitate authorization and secure data management, the proposed framework operates within a broader clinical infrastructure composed of the following core entities:

\begin{itemize}

    \item \textbf{Health Record Repository (HRR):} The HRR is responsible for managing longitudinal electronic health records. It comprises two primary sub-modules:
    
    \begin{itemize}
        \item \textbf{Health Record Controller:} This acts as the central orchestration component. It authenticates users, aggregates contextual evidence from medical IoT devices, hosts the CBAC framework, and manages access request routing and data delivery.
        
        \item \textbf{Health Record Storage:} This serves as the persistent database layer responsible for the secure retention and retrieval of longitudinal patient data. It maintains the structured hierarchy defined in the formal data model, including clinical records $R$, clinical episodes $EP$, and associated provenance metadata such as authorship and subject association.
    \end{itemize}

    \item \textbf{Emergency Authorization Authority (EAA):}
    The EAA is a trusted external entity responsible for validating emergency override requests and issuing EOTs. The EAA verifies the legitimacy of emergency requests based on authenticated requester identity, patient context, and a cryptographic digest of the verified biometric and physiological observation set. Upon successful verification, the EAA issues a signed EOT that binds the authorized semantic disclosure scope, requester, patient identifiers, and a bounded validity interval.

\end{itemize}

\section{Consent Policy Setup and Pre-Commit Validation} \label{sec:consent_policy_setup}

This section formalizes how the proposed CBAC framework maintains a semantically
consistent and analyzable consent repository state across the policy lifecycle,
from consent creation to runtime enforcement. In contrast to traditional access
control systems that defer conflict resolution to request-time evaluation, the
proposed framework enforces semantic correctness at consent commit time, before
patient-issued directives are admitted to the CPR. By shifting validation
upstream, the framework prevents logical inconsistencies and invariant
violations from propagating into the runtime authorization layer.

The section is structured as follows. 

First, Section~\ref{subsec:system_invariants} defines immutable, non-overridable authorization rules that establish baseline access guarantees. Next, Section~\ref{subsec:precommit_validation} describes the pre-commit validation process that governs policy admission, ensuring that only invariant-compliant and conflict-free consent directives transition into an active state. Section~\ref{subsec:consent_evaluation_semantics} then
formalizes the runtime authorization semantics applied by the CPDP over this validated consent state. Finally,
Section~\ref{subsec:consentpolicy_enforcement_workflow} presents the operational enforcement workflow that executes these semantics during standard consent-based access requests.

\subsection{System Invariants (Default Access Policy)}
\label{subsec:system_invariants}

These rules serve as foundational axioms for the CBAC framework.
All subsequent validation and authorization logic is formulated relative to
these axioms to ensure clinical accountability and continuity of care:

\begin{itemize}

    \item \textbf{Creator Access Invariant:}  
    The healthcare professional who authored a clinical episode or record always
    retains access to it.
    \[
    \forall r \in R,\;
    \text{Access}(\mathit{Author}(r), r) = \text{Permit}
    \]

    \item \textbf{Patient Access Invariant:}  
    A patient always retains access to all records that pertain to episodes in
    which they are the subject.
    \[
    \forall r \in R,\;
    \text{Access}(\mathit{Subject}(r), r) = \text{Permit}
    \]

\end{itemize}

\subsection{Pre-Commit Validation Logic}
\label{subsec:precommit_validation}

The consent policy setup workflow governs how patient-issued consent directives
are admitted into the system. During this phase, a patient constructs a draft
consent directive $C_{\text{draft}}$ and submits it to the CPAP. The CPAP performs
preliminary processing and forwards the draft to the CCAM, which acts as an
admission-control mechanism for the consent policy repository.

The CCAM validates each draft directive against two constraints. First, it ensures compliance with the system invariants defined in
Section~\ref{subsec:system_invariants}. Second, it verifies that the directive does not introduce access-mode conflicts with existing active consent directives stored in the CPR.

To enforce invariant compliance, the CCAM resolves the required provenance attributes (e.g., record author and subject) by querying the CPIP, as defined in Section~\ref{subsubsec:clinical_entities}.
In parallel, the CCAM queries the CPR to identify any conflicting active directives applicable to the same grantee and target.

\begin{figure}[htbp]
    \centering
    \includegraphics[width=15cm]{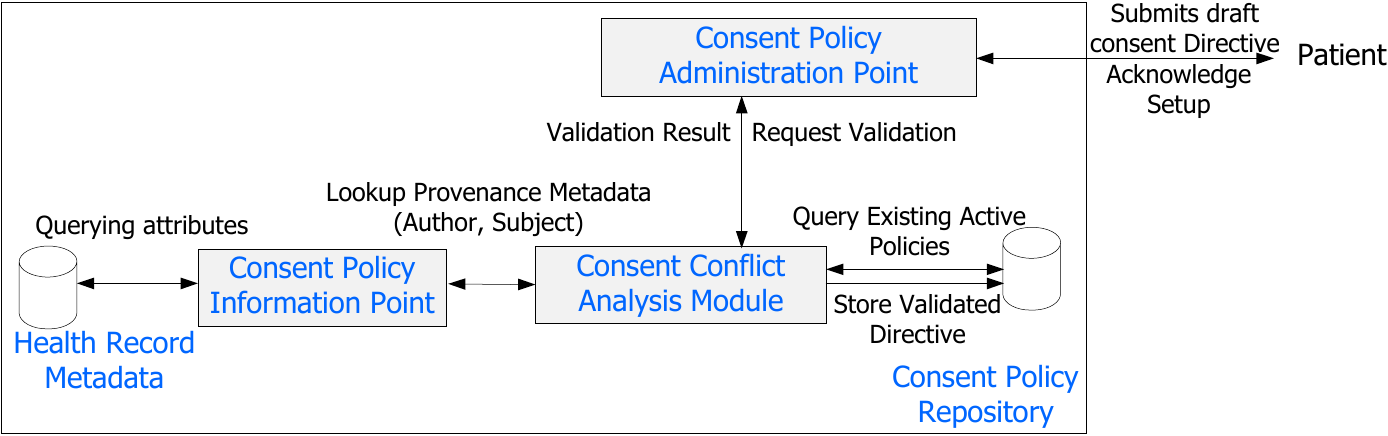}
    \caption{Consent Policy Setup workflow}
    \label{fig:consent_policy_setup}
\end{figure}

Only directives that satisfy both constraints are admitted into the repository
and are transitioned from the \textit{Draft} state to the \textit{Active} state,
becoming eligible for runtime evaluation by the CPDP. 
Figure \ref{fig:consent_policy_setup} illustrates the consent policy setup workflow.

The CCAM identifies two classes of policy anomalies during validation:

\begin{itemize}

\item \textbf{Conflict Type 1: Violation of System Invariants}  \\
A draft directive violates system invariants if it attempts to deny access to
the creator of a clinical episode or record. Formally:
\[
\mathit{Conflict}_{\mathit{inv}}(C_{\mathit{draft}})
\;\Longleftrightarrow\;
\bigl(C_{\mathit{draft}}.\mathit{grantee}
= \mathit{Author}(C_{\mathit{draft}}.\mathit{target})\bigr)
\;\land\;
\bigl(C_{\mathit{draft}}.\mathit{effect} = \text{Deny}\bigr)
\]

\item \textbf{Conflict Type 2: Modality Conflict with Existing Consents}  \\
A modality conflict occurs when a newly submitted directive contradicts an
already active directive by applying an opposing access effect to the same
grantee and target:
\[
\begin{aligned}
\mathit{Conflict}_{\mathit{mod}}(C_{\mathit{draft}})
\;\Longleftrightarrow\;
\exists\, C_{\mathit{exist}} \in \mathcal{C}_{\mathit{active}}
\;\text{s.t.}\;
&\bigl(C_{\mathit{exist}}.\mathit{grantee}
= C_{\mathit{draft}}.\mathit{grantee}\bigr)
\\
&\land\;
\bigl(C_{\mathit{exist}}.\mathit{target}
= C_{\mathit{draft}}.\mathit{target}\bigr)
\\
&\land\;
\bigl(C_{\mathit{exist}}.\mathit{effect}
\neq C_{\mathit{draft}}.\mathit{effect}\bigr)
\end{aligned}
\]

\end{itemize}

\subsection{Consent Evaluation Semantics}
\label{subsec:consent_evaluation_semantics}

For standard consent-based access requests, the CPDP evaluates authorization decisions using a two-layer evaluation model that combines immutable system invariants with patient-delegated consent directives. Since all directives in $\mathcal{C}_{\mathit{active}}$ have passed pre-commit validation, runtime evaluation never encounters invariant violations or modality conflicts.

\begin{enumerate}

\item \textbf{Authorization Layer (System Invariants):}

This layer operationalizes the system invariants defined in
Section~\ref{subsec:system_invariants} during runtime access evaluation. Using
provenance metadata resolved by the CPIP, the CPDP grants immediate authorization
when invariant conditions are satisfied.

\textbf{Logic Evaluation:}
\[
\begin{aligned}
\mathit{Decision}_{\mathrm{Auth}}(sub, r)
\;\Longleftrightarrow\;
&\bigl(sub.\mathit{role} \in \mathit{AllowedRoles}(r)\bigr) \\
&\land\;
\bigl(
sub.\mathit{id} = \mathit{Author}(r)
\;\lor\;
sub.\mathit{id} = \mathit{Subject}(r)
\bigr).
\end{aligned}
\]

\item \textbf{Consent Layer (Patient-Delegated Access):}

If invariant evaluation does not yield an authorization outcome, the CPDP
evaluates patient-delegated consent directives to determine whether access is
explicitly granted beyond baseline guarantees.

\textbf{Logic Evaluation:}
\[
\begin{aligned}
\mathit{Decision}_{\mathit{Consent}}(sub, r, ts)
\;\Longleftrightarrow\;
\exists\, C \in \mathcal{C}_{\mathit{active}}
\;&\text{s.t.}\;
C.\mathit{grantee} = sub.\mathit{id}\\
&\land\; \mathit{InScope}(r, C.\mathit{target})\\
&\land\; \bigl(ts \in C.\mathit{validity}\bigr)\\
&\land\; \bigl(C.\mathit{effect} = \text{Permit}\bigr).
\end{aligned}
\]

\end{enumerate}

\subsection{Consent Policy Enforcement Workflow}
\label{subsec:consentpolicy_enforcement_workflow}

This subsection describes the operational enforcement workflow through which
standard consent-based access requests are processed at runtime. Unlike the
preceding subsections, which define policy constraints and authorization
semantics, this subsection focuses exclusively on how validated consent
directives and system invariants are executed within the CBAC architecture.

When a system user (patient or healthcare professional) initiates an access
request for a health record, the request is intercepted by the C-PEP. The C-PEP normalizes the request into a
standardized tuple, determines the request type, and routes standard
(target-driven) requests to the CPDP for deterministic evaluation.

The CPDP processes each request through the following sequence of operational
steps:

\begin{itemize}

\item \textbf{Attribute Resolution:}
The CPDP queries the CPIP to retrieve the provenance and contextual attributes required for the current request (e.g., author/subject associations, Section~\ref{subsubsec:clinical_entities}).

\item \textbf{Invariant Enforcement:}
Using the resolved attributes, the CPDP enforces the system invariants defined
in Section~\ref{subsec:system_invariants}. If an invariant condition applies,
the CPDP immediately derives a \texttt{Permit} decision without consulting
patient-issued consent directives.

\item \textbf{Consent Directive Retrieval:}
If invariant enforcement does not yield an authorization outcome, the CPDP
queries the CPR to retrieve the set of active consent directives
$\mathcal{C}_{\mathit{active}}$ applicable to the requester. Since all active
directives have passed pre-commit validation, the retrieved set is guaranteed
to be free of invariant violations and modality conflicts.

\item \textbf{Consent Evaluation and Decision:}
The CPDP evaluates the retrieved directives against the request attributes
using the formal consent evaluation semantics defined in
Section~\ref{subsec:consent_evaluation_semantics}. Based on this evaluation, the
CPDP derives a deterministic authorization decision (\texttt{Permit} or
\texttt{Deny}) and returns it to the C-PEP.

\end{itemize}

\begin{figure}[htbp]
    \centering
    \includegraphics[width=15cm]{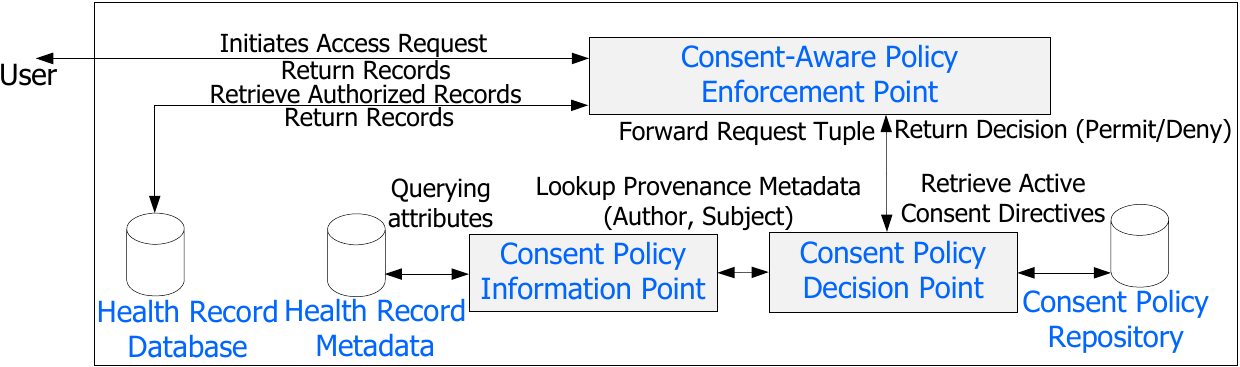}
    \caption{Consent Policy Enforcement workflow}
    \label{fig:consent_policy_enforcement}
\end{figure}

Upon receiving the authorization decision, the C-PEP enforces it by either retrieving and returning the authorized records from the health record database or denying the request. Figure~\ref{fig:consent_policy_enforcement} illustrates the complete enforcement pipeline from request interception to decision
enforcement.

\section{Emergency Override}
\label{sec:emergency_mediator_framework}

This section presents the \emph{Emergency Override Mechanism} of the proposed CBAC framework, which governs controlled \emph{break-the-glass} access to health records when explicit patient consent is unavailable. Emergency override reuses the architectural components introduced in Section~\ref{subsec:architectural_components},
but alters the authorization trigger and disclosure semantics by shifting from
target-driven consent evaluation to evidence-driven, context-aware access authorization.

Section~\ref{subsec:emergency_override_workflow} describes the operational workflow of the emergency override mechanism during an emergency access request, while Section~\ref{subsec:ecdm} formalizes the Emergency Disclosure Control Function (EDCF), which defines the semantic and cryptographic constraints that govern emergency disclosure.

\begin{figure}[htbp]
    \centering
    \includegraphics[width=15cm]{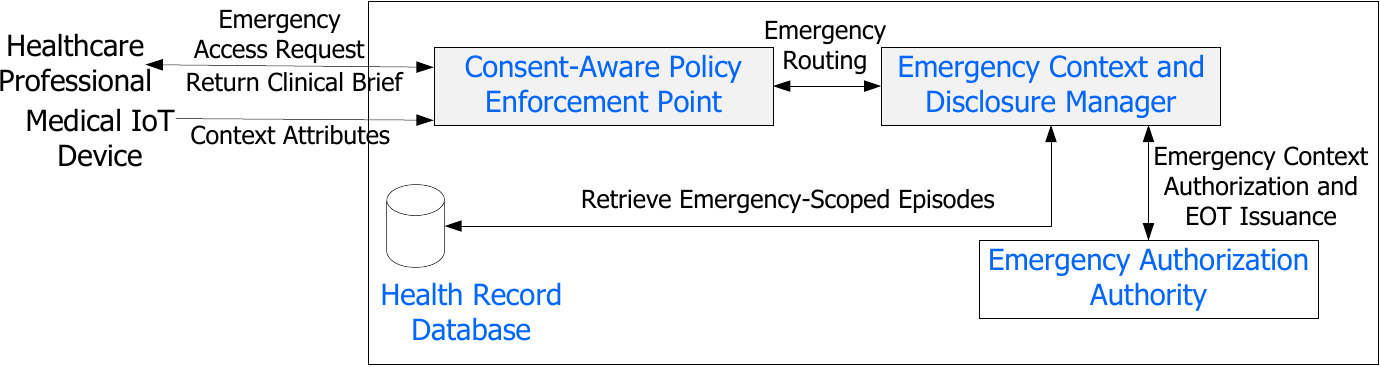}
    \caption{Block diagram of the emergency override authorization processing within the CBAC framework}
    \label{fig:emergencyoverride_blockdiagram}
\end{figure}

\subsection{Emergency Override Workflow}
\label{subsec:emergency_override_workflow}

This subsection presents the end-to-end workflow for emergency override
handling within the proposed CBAC framework during a \emph{break-the-glass}
access request. Because emergency overrides may be triggered without explicit
patient consent, the mechanism must enable timely clinical intervention while
protecting patient privacy through accountability and least-necessary
disclosure. Accordingly, each override is auditable, and minimal data release
requires that emergency access be limited to the smallest set of clinical
records whose semantic labels are relevant to the inferred emergency state~\cite{yang2017lightweight}.

Figure~\ref{fig:emergencyoverride_blockdiagram} presents a block diagram of the
emergency override authorization processing within the CBAC framework, highlighting the roles and interactions of the HRR controller, the ECDM, the EAA, and the health record database during governed break-the-glass access.

Within this workflow, emergency override processing relies on a formally
defined authorization context that consolidates the authenticated requester,
patient identity claim, inferred emergency state, and derived disclosure scope
into a single, verifiable input for downstream authorization. In particular, the
ECDM derives $S_{em}$ and $L_{em}$ from $\mathsf{bioObs}$ and then constructs
$C_{em}$ for submission to the EAA.

\paragraph{Emergency Authorization Context:}
Emergency override requests are evaluated using an abstract emergency
authorization context derived from verified emergency evidence. Formally, the
emergency authorization context is defined as:
\[
C_{em} =
\langle
\mathit{Requester},
\mathit{PatientIDClaim},
S_{em},
L_{em},
ts,
H(\mathsf{bioObs})
\rangle,
\]
where $\mathit{Requester}$ denotes the authenticated healthcare professional in
the logical authorization model, $\mathit{PatientIDClaim}$ represents the
biometric-backed patient identity claim, $S_{em}$ is the inferred set of active
emergency clinical states, $L_{em}$ is the authorized semantic disclosure
scope, $ts$ is the request timestamp, and $H(\mathsf{bioObs})$ is a
cryptographic digest of the verified biometric and physiological observation
set.

The context $C_{em}$ constitutes the sole input to the EAA. It does not itself grant access; rather, it is used to derive a
signed EOT, which encodes the approved disclosure
scope and temporal validity and is subsequently enforced by the HRR
controller.

\begin{enumerate}

\item \textbf{Emergency Authentication and Context Capture}
\begin{itemize}

\item \textbf{Requester authentication:}
The healthcare professional authenticates to the HRR controller using
valid professional credentials. As part of the emergency request initiation, the
HP operates associated medical IoT devices, which capture and transmit
verifiable biometric and physiological observations $\mathsf{bioObs}$ (e.g.,
heart rate, $SpO_2$, blood pressure).

\item \textbf{Patient identification:}
In the absence of explicit patient consent, the HP captures the patient’s
biometric identity. The HRR controller resolves the corresponding
$\mathit{PatientID}$ from this biometric input. The biometric identity
resolution mechanism is treated as a functional precondition and lies
outside the core CBAC authorization logic.

\item \textbf{Emergency request formation:}
The HRR controller aggregates the authenticated requester identity,
resolved $\mathit{PatientID}$, timestamp, and verified physiological evidence
into an emergency access request, which is forwarded to the CBAC framework
and routed to the ECDM.

\end{itemize}

\item \textbf{Emergency Authorization and Scoped Disclosure}
\begin{itemize}

\item \textbf{Emergency context evaluation:}
The ECDM evaluates the physiological evidence $\mathsf{bioObs}$ to infer the active emergency clinical state set $S_{em}$ upon receipt. Using the bipartite relevance model $G$ (see Section~\ref{subsec:ecdm}), the ECDM derives the corresponding semantic emergency disclosure scope $L_{em}$.

\item \textbf{Emergency override authorization:}
The ECDM constructs the emergency authorization context $C_{em}$ and submits
it to the EAA. The EAA verifies the legitimacy of the request based on
authenticated identities, patient context, and the cryptographic digest
$H(\mathsf{bioObs})$. Upon successful verification, the EAA issues a signed EOT that binds the requester identity, patient identity,
authorized semantic scope $L_{em}$, and a bounded validity interval.

\item \textbf{Controlled record access:}
The EOT is returned to the HRR controller, which initiates a scope-bounded
retrieval request to the Health Record Database. Possession and use of the
EOT are logged to ensure post-hoc accountability. The database returns the
set of clinical episodes $EP_{em}$ whose semantic tags intersect with the
authorized scope $L_{em}$.

\item \textbf{Clinical brief generation:}
The HRR controller delivers the authorized episode set $EP_{em}$ to the HP
module, which assembles a condition-relevant Clinical Brief to support
immediate clinical intervention.

\end{itemize}

\end{enumerate}

\begin{figure}[htbp]
    \centering
    \includegraphics[width=11cm]{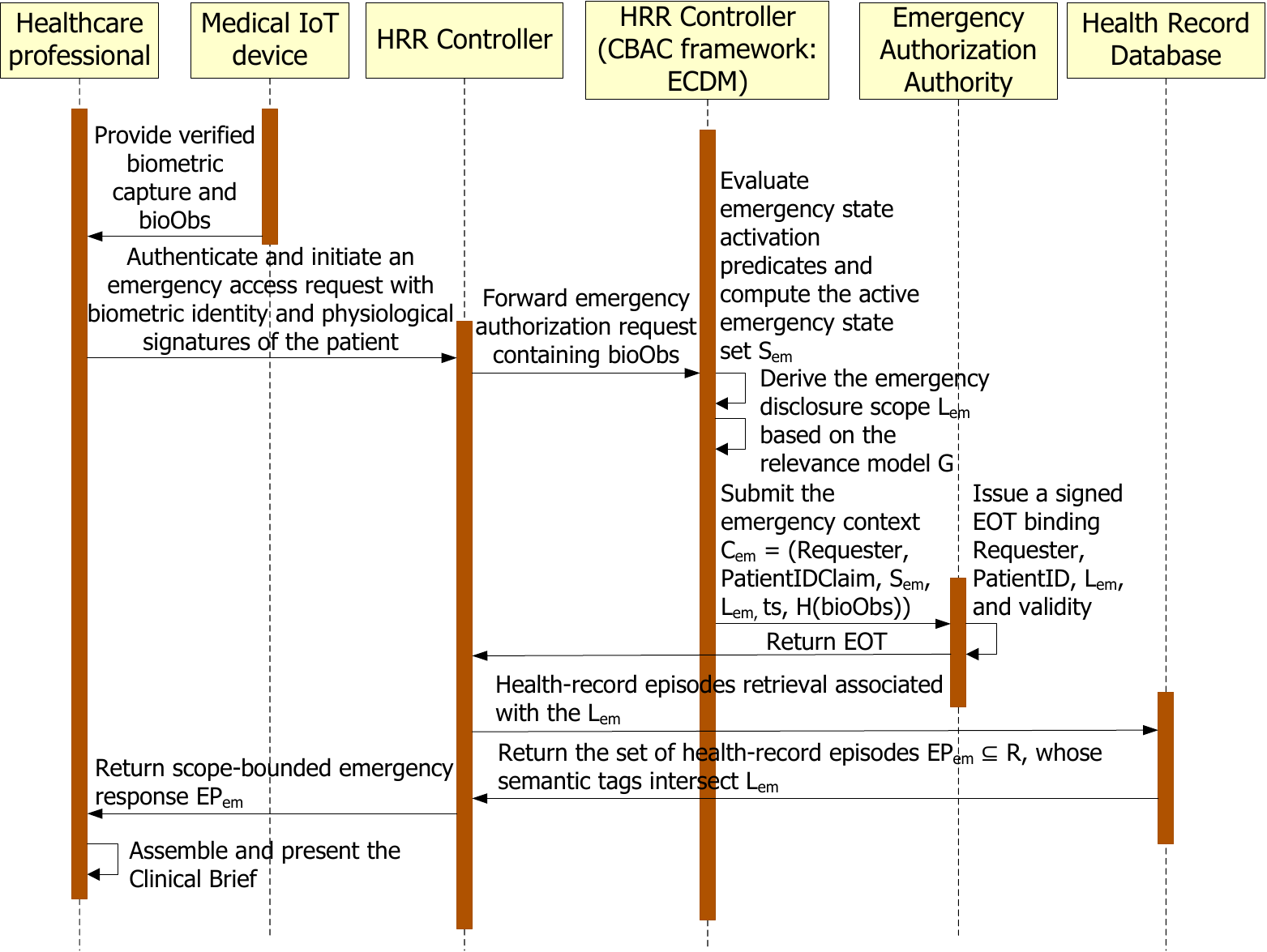}
    \caption{Sequence diagram of context-aware emergency authorization and
    scope-bounded record disclosure during a break-the-glass request}
    \label{fig:emergency_mediator_framework_workflow}
\end{figure}

\subsection{Emergency Context and Disclosure Manager}
\label{subsec:ecdm}

The ECDM is responsible for
deriving a semantically bounded emergency disclosure scope from verified
physiological evidence. Its core function is the EDCF, which maps an authenticated biometric observation set and a
policy-governed relevance model to an authorized disclosure scope:
\[
\textsf{EDCF}(\mathsf{bioObs}, G) = L_{em}
\]
where $L_{em} \subseteq L$ denotes the set of episode-level semantic tags
authorized for emergency access.

\begin{figure}[htbp]
\centering
\includegraphics[width=7cm]{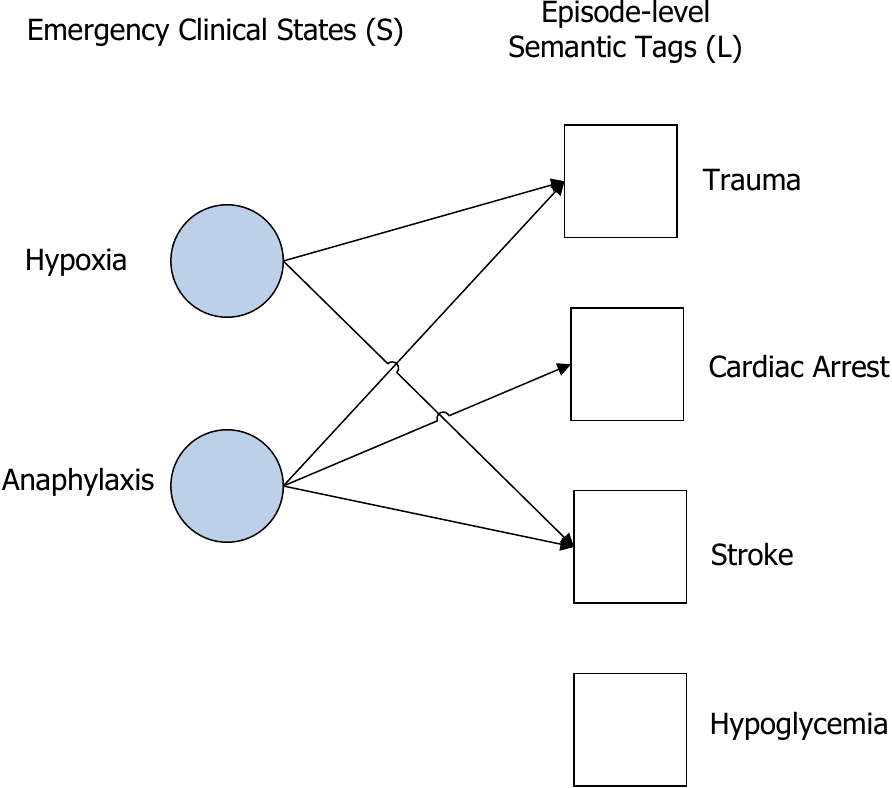}
\caption{Relevance model $G$ linking emergency clinical states $S$ to
episode-level semantic tags $L$}
\label{fig:relevance_model_g}
\end{figure}

The inputs and outputs of the EDCF are defined as follows:

\begin{itemize}

  \item \textbf{Biometric observation set $\mathsf{bioObs}$:} 
  A verifiable set of biometric and physiological observations
  $\{\mathsf{bioObs}_1, \mathsf{bioObs}_2, \dots, \mathsf{bioObs}_n\}$
  collected during an emergency encounter via medical IoT devices. These
  observations constitute authenticated clinical evidence and serve as the
  sole basis for emergency context inference.

  \item \textbf{Relevance model $G$:}
  A static, policy-governed bipartite semantic structure encoding medically
  accepted relevance relationships between emergency clinical states and
  episode-level semantic tags. The model is defined as a tuple:
  \[
  G = (S, L, E),
  \]
  as illustrated in Fig.~\ref{fig:relevance_model_g}, where:
  \begin{itemize}
    \item $S$ is a finite set of emergency clinical states (e.g., hypoxia,
    cardiac arrest).
    \item $L$ is a finite set of episode-level semantic tags (e.g.,
    \textit{Trauma}, \textit{CardiacArrest}, \textit{Stroke}) used to annotate
    health record episodes.
    \item $E \subseteq S \times L$ is a set of directed relevance edges, where
    $(s, l) \in E$ indicates that when the emergency state $s$ is active, episodes
    annotated with the semantic tag $l$ are authorized for consideration during
    emergency access.
  \end{itemize}

\end{itemize}

The output $L_{em}$ represents the minimal semantic disclosure scope derived
from the inferred emergency state set and is subsequently bound into an
EOT by the EAA.
This scope constrains all downstream episode selection and record retrieval
during break-the-glass access.

The internal operational steps of the ECDM, including emergency state
determination, disclosure scope derivation, and EDCF evaluation, are presented
in Section~\ref{subsubsec:ecdm_operational_steps}. A formal characterization of
the safety and minimality properties satisfied by the EDCF is provided in
Section~\ref{subsubsec:ecdm_formal_properties}.

\subsubsection{ECDM Operational Steps}
\label{subsubsec:ecdm_operational_steps}

Once an emergency authorization request is routed to the ECDM, it executes a
deterministic, multi-stage authorization pipeline. The ECDM does not directly
access, retrieve, or assemble clinical data. Instead, it derives a semantic
emergency disclosure scope and coordinates the acquisition of verifiable
emergency override credentials. The pipeline is memoryless and operates solely
on authenticated emergency evidence and policy-defined semantic models.
As illustrated in Fig.~\ref{fig:emergency_mediator_framework_workflow},
emergency override processing is triggered only upon verifiable physiological
evidence and proceeds through a governed authorization path involving the
HRR controller, the CBAC framework, and the EAA.
The internal operation of the ECDM proceeds through the following sequential
stages: (i) Emergency State Determination, (ii) Emergency Disclosure Scope
Derivation, and (iii) Emergency Authorization Token Acquisition.

\begin{enumerate}

\item \textbf{Emergency State Determination}

The Emergency State Determination (ESD) function maps verified biometric and
physiological observations to a well-defined set of active clinical emergency
states. This stage produces a deterministic and auditable semantic abstraction
that grounds all subsequent emergency authorization and disclosure decisions.

\begin{itemize}

\item \textbf{Observation mapping:}
The ECDM evaluates the verified biometric and physiological observation set
$\mathsf{bioObs}$ to infer active emergency clinical states. Since a single
emergency encounter may correspond to multiple concurrent emergencies, the
mapping yields a finite state set:
\[
S_{em} = \mathsf{ESD}(\mathsf{bioObs}),
\]
where $S_{em} \subseteq S$ denotes the set of emergency clinical states inferred
from the observation set.

\item \textbf{Emergency state activation predicates:}
Each emergency state $s \in S$ is associated with a deterministic activation
predicate:
\[
\textsf{EMStateAct}_s : \mathsf{bioObs} \rightarrow \{0,1\},
\]
which evaluates whether the observed physiological evidence satisfies the
clinical criteria for activating state $s$.

\item \textbf{State activation rule:}
The set of active emergency states is defined as:
\[
S_{em} =
\{\, s \in S \mid \textsf{EMStateAct}_s(\mathsf{bioObs}) = 1 \,\}.
\]

\end{itemize}

\item \textbf{Emergency Disclosure Scope Derivation}

Given the inferred emergency state set $S_{em}$, the ECDM derives a semantic
emergency disclosure scope consisting of episode-level clinical tags. Each tag
$l \in L$ corresponds to a clinical category authorized for emergency access
under the relevance model $G$:
\[
L_{em}(S_{em}) =
\{\, l \in L \mid \exists s \in S_{em} \text{ such that } (s,l) \in E \,\}.
\]
The derived set $L_{em}$ defines the minimal semantic scope permitted for
emergency disclosure under the current emergency context.

\item \textbf{Emergency Authorization Token Acquisition}

Following the derivation of the emergency disclosure scope $L_{em}$, the ECDM
constructs an emergency authorization request and submits it to the EAA. The request binds the authenticated requester
identity, the resolved patient context, a cryptographic hash of the verified
biometric and physiological observation set $\mathsf{bioObs}$, and the derived
semantic scope $L_{em}$.

Upon successful verification, the EAA issues a signed EOT that cryptographically attests to the legitimacy of the emergency request.
The EOT encodes the authorized semantic disclosure scope, requester identity,
patient identifier, and a bounded validity interval. The issued token serves as
a verifiable authorization credential that enables the HRR controller to
retrieve only those health records whose semantic annotations fall within the
authorized emergency scope.

\end{enumerate}

\subsubsection{ECDM Formal Properties}
\label{subsubsec:ecdm_formal_properties}

This subsection formalizes the key properties satisfied by the EDCF, which is executed internally by the ECDM.
Collectively, these properties ensure bounded disclosure, clinical safety, and
predictable behavior under emergency override conditions.

\begin{enumerate}

\item \textbf{Determinism}

\begin{itemize}
    \item \textbf{Statement:}
    For a fixed relevance model $G$, the EDCF is deterministic. For each biometric observation set, it derives a
    single emergency disclosure scope, denoted by:
    \[
    L_{em} \;\triangleq\; \textsf{EDCF}(\mathsf{bioObs}, G)
    \]

    \item \textbf{Rationale:}
    Each stage of the EDCF pipeline is deterministic.
    The mapping from verified biometric observations to emergency clinical
    states, and from emergency states to the semantic disclosure scope,
    relies exclusively on static activation predicates and a fixed relevance
    model.
    As a result, the EDCF generates a unique and reproducible semantic
    authorization outcome.
\end{itemize}

\item \textbf{Bounded Disclosure}

\begin{itemize}
    \item \textbf{Statement:}
    For any emergency access authorized under an issued EOT, the semantic disclosure scope $L_{em}$ encoded in the token strictly bounds the
    episodes that can be retrieved. Let $EP_{em} = \mathsf{retrieve}(\mathit{EOT})$ denote the disclosed episode set, then:
    \[
    \forall e \in EP_{em},\;
    \textsf{semTag}(e) \cap L_{em} \neq \emptyset
    \]

    \item \textbf{Rationale:}
    The EDCF derives $L_{em}$ exclusively from the active emergency state set
    via the relevance model $G$. During emergency access, the HRR controller
    uses the issued EOT, which cryptographically
    binds $L_{em}$, to constrain record retrieval. No clinical episode whose
    semantic annotations fall outside $L_{em}$ can be retrieved or disclosed.
    Possession and use of the EOT are logged to support post-hoc accountability.
\end{itemize}

\item \textbf{Consent Non-Interference}

\begin{itemize}
    \item \textbf{Statement:}
    The emergency disclosure scope derived by the EDCF is independent of patient-defined consent directives. In particular, for a fixed $\mathsf{bioObs}$ and $G$, the derived scope is invariant under any change to the consent repository state:
    \[
    \textsf{EDCF}(\mathsf{bioObs}, G)
    \quad\text{is independent of }\mathcal{C}_{\mathit{active}}
    \]

    \item \textbf{Rationale:}
    Emergency override authorization is governed solely by verified emergency
    observations, the relevance model $G$, and clinical safety requirements.
    Patient-defined consent directives are not consulted during emergency
    scope derivation. As a result, restrictive, expired, or absent consent
    directives cannot obstruct safety-critical access during emergency
    situations.
\end{itemize}

\item \textbf{Traceability}

\begin{itemize}

\item \textbf{Statement:}
Emergency access is authorized through EOTs. For every clinical episode disclosed during emergency access, there exists an EOT that justifies and constrains the disclosure to an authorized semantic scope:
\[
\forall e \in EP_{em},\;
\exists\, \mathit{EOT}
\text{ such that }
e \in \mathsf{retrieve}(\mathit{EOT})
\]

The inferred emergency clinical state $S_{em}$ and the relevance model $G$ determine the authorized semantic disclosure scope $L_{em}$ encoded in the EOT. A disclosed episode is semantically sound only if its annotations intersect their authorized scope:
\[
\forall e \in EP_{em},\;
\textsf{semTag}(e) \cap L_{em} \neq \emptyset
\]

\item \textbf{Rationale:}
This property ensures that every emergency disclosure is attributable to a
cryptographically issued EOT that binds the requester
identity, patient identity, and authorized semantic scope. Because record
retrieval is strictly constrained by the semantic scope encoded in the EOT, each
disclosure is explainable and auditable. This enables robust post-hoc
accountability while preventing opportunistic data disclosure during break-the-glass access.

\end{itemize}

\end{enumerate}

\section{Performance Measurement} \label{sec:performance_measurement}

This section evaluates the performance of the proposed CBAC framework through a series of controlled simulations. The experiments focus on three key aspects: (i) the computational overhead of pre-commit consent validation (Section~\ref{subsec:perf-eval-precommit-consent-validation}), (ii) the impact of proactive anomaly pruning on runtime policy decision efficiency (Section~\ref{subsec:impact-precommit-runtime-enforcement}), and (iii) the effectiveness of context-aware minimal data release in emergency \emph{break-the-glass} scenarios (Section~\ref{subsec:evaluation-emergency-context-aware-minimal-release}).

All simulations were conducted on a uniform hardware platform (Intel Core i5-1235U processor with 16 GB of RAM) using Python-based custom simulators. To ensure a fair comparison, both the standard XACML PDP and the proposed CPDP rely on the same underlying policy evaluation engine, AuthzForce~\cite{authzforce}, an open-source Java implementation of the OASIS XACML~3.0 standard.

Both the baseline XACML system and the proposed CBAC framework use the same PDP implementation (AuthzForce) across all simulations, including the same unmodified mechanisms for policy storage, retrieval, and evaluation. Therefore, any general performance improvements to AuthzForce’s internal policy management (e.g., caching or indexing) would be expected to benefit both approaches. The runtime decision-latency gains reported in Section~\ref{subsec:impact-precommit-runtime-enforcement} result from resolving conflicts and redundancies at consent creation time, which reduces the effective policy set that must be evaluated during runtime authorization.

All simulation scripts and custom datasets used in this evaluation are publicly available at \texttt{github.com/nasif2005/cbac-precommit-emergency-eval}.~\cite{nasif2005_cbac_precommit_emergency_eval}

Each of the following subsections adheres to a consistent experimental structure:
\begin{itemize}
  \item \textbf{Dataset generation}: Description of the synthetic policy repositories or health-record metadata used, including size and controlled anomaly injection rates.
  \item \textbf{Workload}: Definition of the input stream (consent directives or access requests) or emergency scenarios applied during the experiment.
  \item \textbf{Simulation Results and Analysis}: Presentation and interpretation of measured performance metrics (processing time, disclosure ratio) and their implications for the CBAC framework.
\end{itemize}

\subsection{Performance Evaluation of Pre-Commit Consent Validation} \label{subsec:perf-eval-precommit-consent-validation}

The objective of this simulation-based evaluation is to quantify the computational overhead of the pre-commit validation workflow, specifically measuring the latency introduced by the CCAM during the policy setup phase.

\begin{figure}[htbp]
        \centering
        \includegraphics[width=8cm]{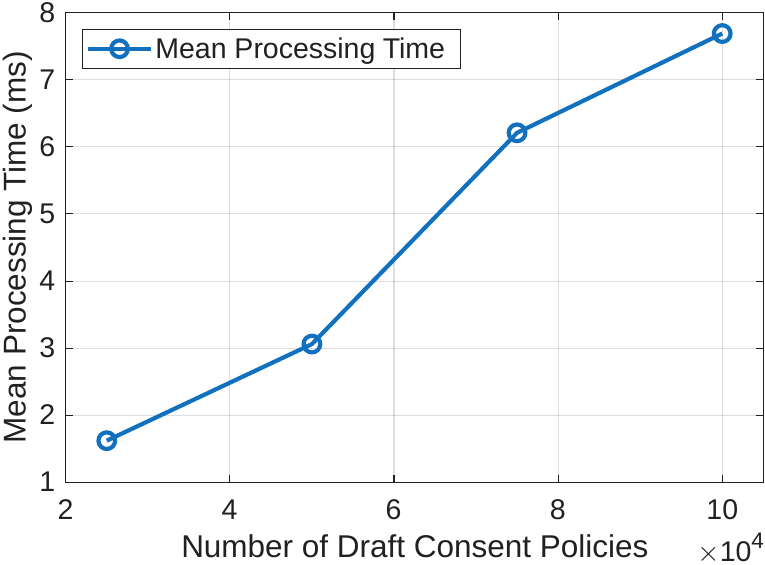}
        \caption{Mean processing time during pre-commit consent policy setup}
        \label{fig:consent_policy_setup_processingtime}
    \end{figure}

\begin{itemize}
    
    \item \textbf{Dataset generation}:
    To evaluate framework performance under a controlled increase in the number of policies within the CPR, a synthetic health-record metadata dataset was generated, consisting of $E = 100000$ episodes and an input stream of $N = 100000$ draft consent directives.
    To model sustained growth under a fixed anomaly density, the input stream was seeded with an overall anomaly rate of $20\%$, comprising $10\%$ modality conflicts and $10\%$ redundant directives.
    The stream was then deterministically shuffled to ensure a uniform distribution of anomalies across all subsets.

    \item \textbf{Workload}:
    The proposed CBAC framework’s scalability was evaluated by processing the input stream in incremental subsets of size $N \in {25000, 50000, 75000, 100000}$.
    For each directive in a subset, the framework performed (i) syntactic validation and (ii) semantic validation against the current state of the active repository to identify and prune modality conflicts and redundancies prior to commitment.

    \item \textbf{Simulation Results and Analysis}: 
    The simulation results demonstrate that setup latency increases monotonically as the repository scales. Specifically, the mean per-consent setup time rises from 1.62 ms at $N=25000$ to 7.69 ms at $N=100000$. This trend is attributable to the increasing complexity of semantic validation, as the CCAM must verify each draft consent directive against a larger set of existing active consent policies to detect modality conflicts and redundancies. 
    While this validation introduces measurable latency during the setup phase, it represents a one-time computational investment. By shifting the burden of conflict resolution from runtime enforcement to the pre-commit phase, the framework ensures that the policy repository remains in a consistent, anomaly-free state.     
    Figure~\ref{fig:consent_policy_setup_processingtime} shows that the per-consent validation cost remains within single-digit milliseconds even at the largest evaluated scale, confirming that pre-commit validation introduces a predictable and bounded overhead.

\end{itemize}

\subsection{Impact of Pre-Commit Validation on Runtime Enforcement Efficiency}
\label{subsec:impact-precommit-runtime-enforcement}

The objective of this simulation-based evaluation is to evaluate how the proactive pruning of consent policy anomalies during the setup phase improves the performance of the CPDP during runtime enforcement.

\begin{figure}[t]
      \centering
    
      \begin{subfigure}{0.48\linewidth}
        \centering
        \includegraphics[width=\linewidth]{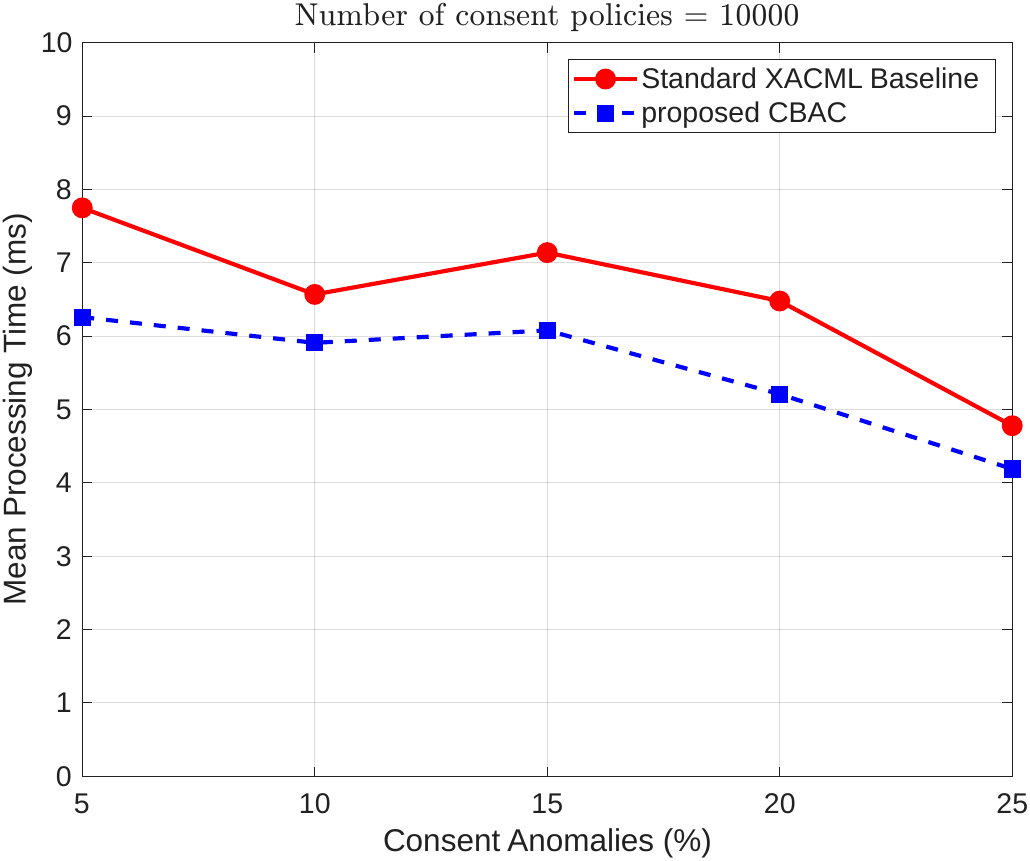}
        \caption{10000 consent policies}
      \end{subfigure}
      \hfill
      \begin{subfigure}{0.48\linewidth}
        \centering
        \includegraphics[width=\linewidth]{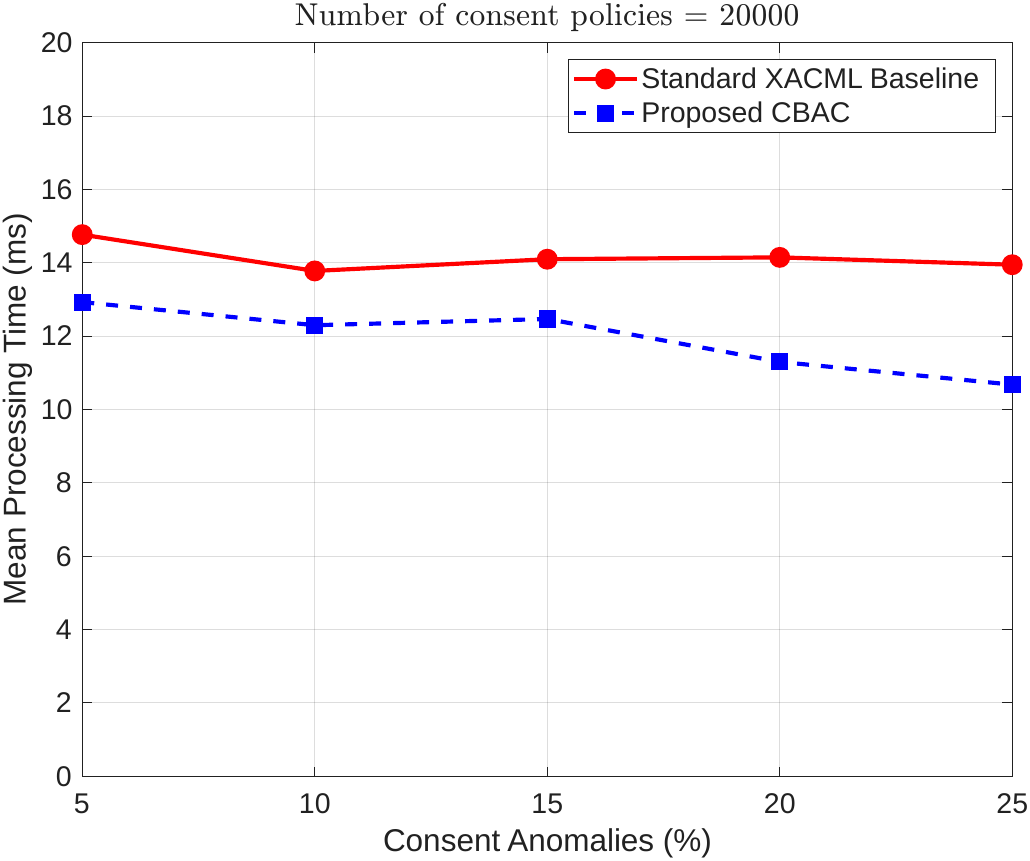}
        \caption{20000 policies}
      \end{subfigure}
    
      \vspace{0.8em}
    
      \begin{subfigure}{0.48\linewidth}
        \centering
        \includegraphics[width=\linewidth]{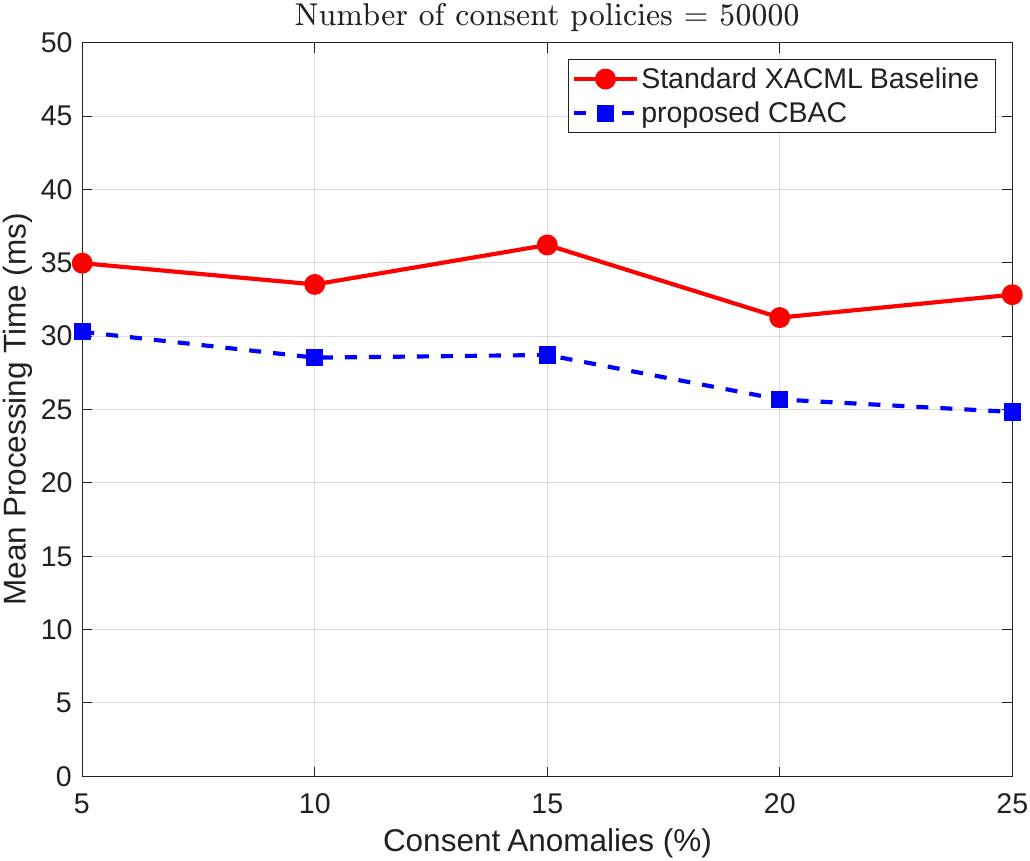}
        \caption{50000 policies}
      \end{subfigure}
      \hfill
      \begin{subfigure}{0.48\linewidth}
        \centering
        \includegraphics[width=\linewidth]{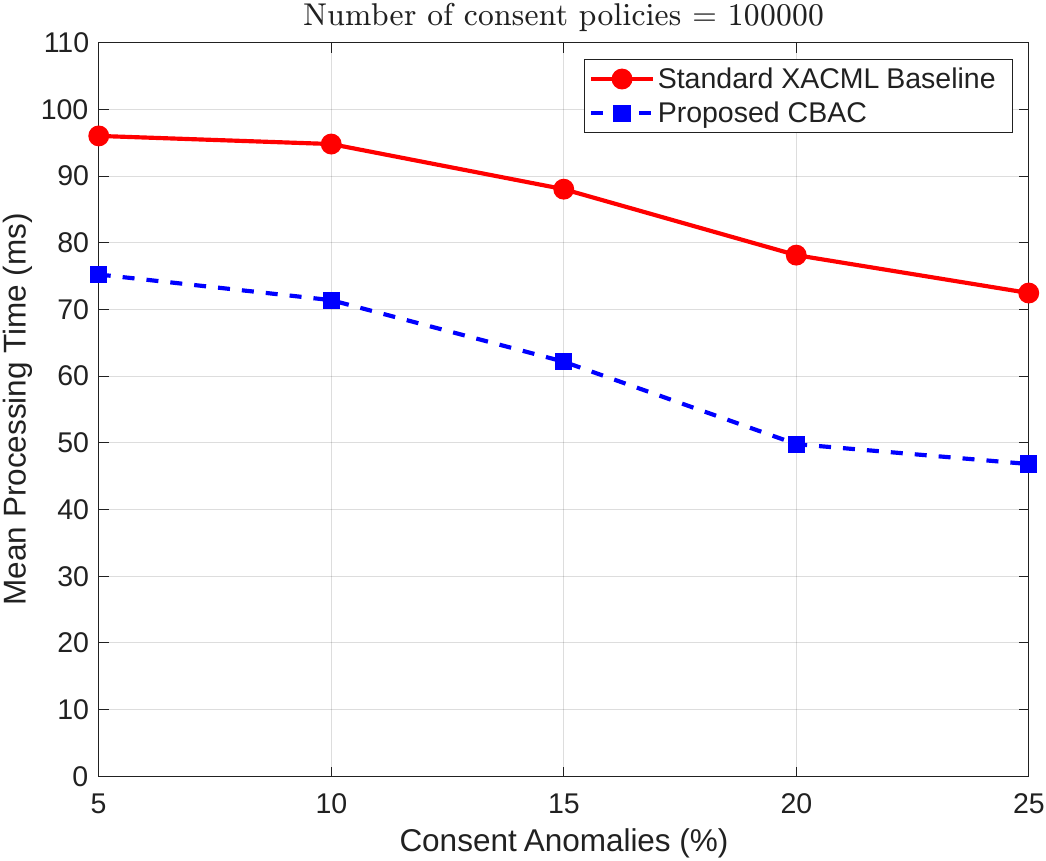}
        \caption{100000 policies}
      \end{subfigure}
    
      \caption{Mean request processing time across policy repository sizes}
      \label{fig:mean-processing-time}
    \end{figure}

\begin{itemize}

  \item \textbf{Dataset generation:}
   Synthetic consent policy repositories were generated with fixed sizes 
   $N \in \{10000, 20000, 50000, 100000\}$ and anomaly injection rates ranging from 5\% to 25\%. 
   For each combination of repository size and anomaly rate, an independently generated policy set was constructed. 

  \item \textbf{Workload:}
  A standardized workload $Q$ consisting of 1000 access requests (60\% applicable requests and 40\% non-applicable requests) was reused across all experimental trials to ensure comparative consistency.

  \item \textbf{Simulation Results and Analysis:}
   The proposed CBAC framework consistently outperforms the standard XACML baseline in runtime decision latency across all evaluated repository sizes and anomaly rates (Figure~\ref{fig:mean-processing-time}). This advantage arises from pre-commit anomaly pruning, which proactively removes contradictory and redundant directives, thereby reducing the effective number of policies evaluated per request.

   For a repository of 10000 policies (Figure~\ref{fig:mean-processing-time}(a)), the baseline exhibits higher latency (7.0--7.8 ms) and noticeable variability, peaking near 15\% anomalies. In contrast, the proposed framework achieves significantly lower latency (4.2--6.3 ms), which decreases monotonically with increasing anomaly rate. This trend reflects greater pruning efficacy at higher anomaly densities, resulting in progressively smaller active policy sets.

   The performance gap widens at larger scales (20000 to 100000 policies; Figure~\ref{fig:mean-processing-time}(b--d)), where CBAC maintains sub-7 ms decisions while the baseline exceeds 10--15 ms in several configurations. These results confirm that shifting anomaly resolution to the infrequent pre-commit phase yields substantial runtime gains in healthcare environments, where policy updates are rare compared to high-frequency access requests.
  
\end{itemize}

\subsection{Evaluation of Emergency Context-Aware Minimal Data Release}
\label{subsec:evaluation-emergency-context-aware-minimal-release}

The objective of this simulation-based evaluation is to assess the effectiveness of the EDCF in balancing clinical necessity with patient privacy during \emph{break-the-glass} scenarios.

\begin{itemize}
   
   \item \textbf{Dataset generation:} Synthetic health records were generated for two distinct repository sizes, $R=500$ and $R=1000$. Each record element was associated with one or more semantic labels corresponding to the categories defined in the relevance model $G$.
   
   \item \textbf{Workload:} The simulation evaluated the EDCF across six clinical emergency states: Trauma, Cardiac Arrest, Hypoxia, Anaphylaxis, Stroke, and Hypoglycemia. For each scenario, the ECDM mapped verified biometric and physiological observations $\mathsf{bioObs}$ to the corresponding emergency clinical states and derived the authorized emergency disclosure scope $L_{em}$.

   \item \textbf{Simulation Results and Analysis:}
   The privacy effectiveness of the EDCF is evaluated using the Disclosure Ratio $\eta = |D| / |R|$, representing the proportion of the total record $R$ included in the condition-relevant Clinical Brief. Tables~\ref{tab:privacy_effectiveness_500} and~\ref{tab:privacy_effectiveness_1000} show that the EDCF effectively prunes non-essential elements while retaining life-saving information tied to biometric triggers.
   For the $|R| = 500$ dataset (Table~\ref{tab:privacy_effectiveness_500}), $\eta$ ranges from 0.220 (Trauma) to 0.070 (Hypoglycemia), achieving 78.0--93.0\% pruning.
   At $|R| = 1000$ (Table~\ref{tab:privacy_effectiveness_1000}), $\eta$ decreases further (e.g., to 0.045 for Hypoglycemia), reflecting enhanced relative privacy as the total record size grows while relevant elements remain bounded by the semantic scope of each emergency state in $G$. On average, pruning improves by $\sim$5--7\% relative to the smaller repository.
   These findings validate the EDCF's ability to enforce context-aware, minimal disclosure during break-the-glass access, ensuring scalability and post-hoc accountability in safety-critical healthcare settings. While pruning ratios may vary in real-world deployments using actual clinical records, these results demonstrate the relative effectiveness and consistent scaling behavior of the EDCF under controlled conditions.
   
\end{itemize}

   % \paragraph{Privacy Effectiveness: The Minimization Ratio}
    
    % --- Simulation Output: Privacy Effectiveness (|R| = 500) ---
    \begin{table}[ht]
    \centering
    \caption{Simulation Output: Privacy Effectiveness ($|R| = 500$)}
    \label{tab:privacy_effectiveness_500}
    \begin{tabular}{|l|l|c|c|c|c|}
    \hline
    \textbf{Case} & \textbf{Emergency State} & $|R|$ & $|D|$ & \textbf{Disclosure Ratio ($\eta$)} & \textbf{Pruned (\%)} \\
    \hline
    E01 & Trauma         & 500 & 110 & 0.220 & 78.0\% \\
    \hline
    E02 & Cardiac Arrest & 500 & 84  & 0.168 & 83.2\% \\
    \hline
    E03 & Hypoxia        & 500 & 100 & 0.200 & 80.0\% \\
    \hline
    E04 & Anaphylaxis    & 500 & 73  & 0.146 & 85.4\% \\
    \hline
    E05 & Stroke         & 500 & 85  & 0.170 & 83.0\% \\
    \hline
    E10 & Hypoglycemia   & 500 & 35  & 0.070 & 93.0\% \\
    \hline
    \end{tabular}
    \end{table}
    
    % --- Simulation Output: Privacy Effectiveness (|R| = 1000) ---
    \begin{table}[ht]
    \centering
    \caption{Simulation Output: Privacy Effectiveness ($|R| = 1000$)}
    \label{tab:privacy_effectiveness_1000}
    \begin{tabular}{|l|l|c|c|c|c|}
    \hline
    \textbf{Case} & \textbf{Emergency State} & $|R|$ & $|D|$ & \textbf{Disclosure Ratio ($\eta$)} & \textbf{Pruned (\%)} \\
    \hline
    E01 & Trauma         & 1000 & 145 & 0.145 & 85.5\% \\
    \hline
    E02 & Cardiac Arrest & 1000 & 116 & 0.116 & 88.4\% \\
    \hline
    E03 & Hypoxia        & 1000 & 146 & 0.146 & 85.4\% \\
    \hline
    E04 & Anaphylaxis    & 1000 & 89  & 0.089 & 91.1\% \\
    \hline
    E05 & Stroke         & 1000 & 114 & 0.114 & 88.6\% \\
    \hline
    E10 & Hypoglycemia   & 1000 & 45  & 0.045 & 95.5\% \\
    \hline
    \end{tabular}
    \end{table}

\section{Discussion, Conclusion, and Future Work}\label{sec:discussion}

This work reconsiders a foundational design assumption in consent-based access control for healthcare systems: that semantic correctness and conflict resolution should be deferred to runtime policy evaluation. In traditional XACML-based architectures, overlapping or contradictory consent directives are resolved lazily through policy-combining algorithms when an access request is evaluated. While this approach preserves expressive flexibility, it shifts assurance away from the policy repository itself and toward request-dependent arbitration. In safety-critical clinical environments, this trade-off complicates predictability, explainability, and performance under high request volumes.

The proposed framework addresses these limitations by elevating semantic validation to a mandatory pre-commit phase during consent setup. By detecting modality conflicts, invariant violations, and structural inconsistencies before directives become active, the system maintains a semantically consistent policy repository. Authorization decisions are therefore derived from a validated and stable policy state, rather than from ad hoc resolutions among conflicting rules at runtime. This architectural shift relocates semantic reasoning to the phase when human intent is expressed and repository-wide context is available, while simplifying the runtime responsibilities of the CPDP. As demonstrated in the evaluation, this design reduces runtime decision latency and eliminates request-order sensitivity, yielding predictable and explainable authorization outcomes under realistic clinical workloads.

Beyond performance, the framework strengthens safety and continuity of care through two complementary mechanisms. First, formally defined system invariants guarantee immutable baseline access for record authors and patients themselves, preventing overly restrictive or contradictory consent directives from unintentionally disrupting clinical workflows. Second, emergency access is handled through a controlled override mechanism within the authorization logic, enabling time-critical disclosure when usual consent constraints would otherwise deny access. Rather than granting unrestricted break-the-glass privileges, emergency disclosures are constrained to condition-relevant data and are justified by contextual evidence, ensuring proportionality, auditability, and post-event accountability. Together, these mechanisms demonstrate that enforcing semantic correctness at consent creation time can preserve patient autonomy without compromising clinical urgency or operational efficiency.

The proposed framework operates under several deliberate design assumptions. It assumes a trusted centralized infrastructure, particularly the HRR, reflecting prevailing healthcare deployment models in which regulated institutions operate authorization and data management. While this assumption enables efficient policy evaluation and emergency handling, enforcement correctness ultimately depends on the integrity of these components. In addition, the relevance model used to constrain emergency disclosure is static and policy-defined, prioritizing predictability and auditability over adaptive behavior. Finally, consent enforcement in the current design is logical rather than cryptographic, relying on trusted execution rather than the cryptographic enforcement of access constraints.

These limitations motivate several directions for future work. 
A primary avenue is the integration of cryptographic enforcement mechanisms that reduce reliance on trusted intermediaries while preserving the semantic guarantees established in this work. Embedding consent semantics directly into cryptographic access mechanisms would enable verifiable enforcement even in the presence of partially compromised infrastructure.

A second direction involves extending the emergency relevance model to support
carefully bounded adaptability. The relevance model is currently static; future
work could support guideline-driven updates while preserving traceability.

Finally, the framework provides a natural foundation for integrating
LLM-assisted consent authoring interfaces. By coupling probabilistic natural
language interpretation with deterministic pre-commit semantic validation,
future systems could allow patients to express nuanced consent preferences in
natural language while ensuring that only conflict-free, semantically consistent
directives are admitted into the policy repository. This combination offers a
promising path toward usable, safe, and explainable consent management in
complex healthcare environments.

In summary, this work demonstrates that proactive semantic validation,
combined with invariant-aware authorization and context-constrained emergency
override, provides a principled and practical foundation for consent-based
access control in healthcare systems.

\bibliographystyle{IEEEtran}
\bibliography{IEEEfull}

\end{document}